# Atomic layer deposition of HfO$_2$ on graphene from HfCl$_4$ and H$_2$O

Harry Alles[1,2], Jaan Aarik[1], Aleks Aidla[1], Aurelien Fay[2], Jekaterina Kozlova[1,3], Ahti Niilisk[1], Martti Pärs[1,*], Mihkel Rähn[1], Maciej Wiesner[2,4], Pertti Hakonen[2], and Väino Sammelselg[1,3]

[1]Institute of Physics, University of Tartu, Tartu, 51014, Estonia

[2]Low Temperature Laboratory, Aalto University, Espoo, P.O. Box 15100, Finland

[3]Institute of Chemistry, University of Tartu, Tartu, 50011, Estonia

[4]Faculty of Physics, Adam Mickiewicz University, Poznan, 61-614, Poland

**Abstract**

Atomic layer deposition of ultrathin HfO$_2$ on unmodified graphene from HfCl$_4$ and H$_2$O was investigated. Surface RMS roughness down to 0.5 nm was obtained for amorphous, 30 nm thick hafnia film grown at 180ºC. HfO$_2$ was deposited also in a two-step temperature process where the initial growth of about 1 nm at 170 ºC was continued up to 10–30 nm at 300 ºC. This process yielded uniform, monoclinic HfO$_2$ films with RMS roughness of 1.7 nm for 10–12 nm thick films and 2.5 nm for 30 nm thick films. Raman spectroscopy studies revealed that the deposition process caused compressive biaxial strain in graphene whereas no extra defects were generated. An 11 nm thick HfO$_2$ film deposited onto bilayer graphene reduced the electron mobility by less than 10% at the Dirac point and by 30–40% far away from it.

# 1. Introduction

Graphene, a single sheet of hexagonally bonded carbon atoms, is currently counted as one of the most promising materials that could allow the Moore's law to be tracked for several more years after the silicon road map ends. This is because graphene has high mobility of charge carriers at room temperature as it was demonstrated already in the very first experiments in 2004 [1]. Electronic transport properties of graphene have been investigated mostly for uncoated single- or few-layer sheets prepared by micromechanical cleaving of graphite or by decomposition of silicon carbide [2]. However, in order to develop real graphene-based electronic devices, one has to be able to prepare efficient top gates with ultra-thin high-k dielectrics on graphene. Recent progress in the preparation of wafer-scale graphene either synthesized by chemical vapor deposition [3–5] or by thermal treatment of SiC [2,6] has made this task even more important.

Atomic layer deposition (ALD) is probably the most suitable method for controlled deposition of ultrathin dielectric films with uniform thickness [7]. Unfortunately, the initiation of ALD on graphene is difficult due to the lack of dangling bonds on the graphene plane, which is similar to the surface of carbon nanotubes [8]. In the case of carbon nanotubes, the surface has been functionalized using nitrogen dioxide ($NO_2$) and trimethylaluminum ($Al[CH_3]_3$), and subsequently coated with $Al_2O_3$ by ALD [9]. Similarly $Al_2O_3$ has been deposited also on graphene [10].

Functionalization, however, can lead to severe degradation of transport properties of graphene [11]. For this reason there have been attempts to grow dielectrics (e.g. $Al_2O_3$ and $HfO_2$) directly on as-cleaved graphene and HOPG [12–14]. In most cases conventional $H_2O$-based ALD processes did not yield uniform metal oxide layers on these surfaces. Nevertheless, Meric *et al.* [15] were recently able to deposit $HfO_2$ directly on graphene using $[(CH_3)_2N]_4Hf$ and $H_2O$ as precursors. According to them the growth was most likely due to physisorption of a precursor on graphene at a very low substrate temperature (90 ºC). Unfortunately the surface of a $HfO_2$ film was noticeably rougher on graphene than on $SiO_2$. In addition, $HfO_2$ films, grown by ALD at low temperatures, have relatively small dielectric constants [16,17]. Thus, in order to obtain gate dielectrics with higher dielectric constant, it would be desirable to increase the deposition temperature.

In this paper we report on ALD of $HfO_2$ films on non-functionalized graphene surface grown from $HfCl_4$ and $H_2O$ as precursors. By using a two-step process in which a $HfO_2$ seed layer was deposited at low temperature and the rest of the layer at high temperature, we succeeded to grow monoclinic $HfO_2$ films on graphene without producing additional defects in graphene. The significant compressive strain, which was detected in a single-layer graphene by Raman spectroscopy, indicated strong adhesion of $HfO_2$ to graphene.

## 2. Experimental

Our graphene samples were micromechanically extracted from natural (Madagascar) graphite and transferred onto Si substrates covered by 250 nm thick $SiO_2$ layer. The as-cleaved graphene flakes were characterized using Raman spectroscopy and atomic force microscopy (AFM) methods in order to determine the number and quality of graphene layers in those. For further experiments, samples with single- and bilayer graphene flakes were selected on the basis of these studies.

Onto one bilayer graphene sample on a highly doped Si substrate, we prepared Ti/Al/Ti electrodes with 10/50/5 nm layer thicknesses in order to perform electrical measurements. Six 0.3 µm wide electrodes with the distance of 0.3 µm between them were formed on a graphene stripe with the width of 1.3 µm using e-beam lift-off lithography and the electron beam evaporation methods. The bottom and top resists were made from polymethyl methacrylate and methacrylic acid (PMMA/MAA) diluted in ethyl-lactate (10%) and PMMA diluted in anisole (3%), respectively. Both resist layers wer baked at 180 ºC for 5 min. Development was made in the solution of methyl isobutyl ketone and isopropyl alcohol (MIBK/IPA) for 10 s and in IPA for 20 s. After deposition of the metal layers the lift-off was done in a warm acetone.

The deposition of $HfO_2$ was studied in more detail on as-cleaved graphene. The $HfO_2$ films were deposited from $HfCl_4$ and $H_2O$ in a flow-type low-pressure ALD reactor [18]. In order to synthesize $HfO_2$, an ALD cycle consisting of an $HfCl_4$ pulse (5 s in duration), purge of the reaction zone with $N_2$ (2 s), $H_2O$ pulse (2 s) and another purge (5 s) was repeated until a film of required thickness was obtained. $HfCl_4$ was volatilized at a temperature of 140 ºC in the flow of $N_2$ carrier gas. The $HfCl_4$ pulses were formed by changing the flow direction of the carrier gas between the $HfCl_4$ source and reaction zone. The $H_2O$ source was kept at the room temperature while the $H_2O$ vapour was led to the reactor through a needle and solenoid valves, which controlled the precursor supply. At the reactor outlet, the partial pressure of $H_2O$ was 5 Pa during an $H_2O$ pulse while the $N_2$ pressure was approximately 250 Pa during the whole deposition process. The choice of precursor pressures as well as of pulse and purge durations was based on earlier real-time quartz crystal microbalance measurements [18,19] that demonstrated reliable self-limited character of the deposition at the process parameters used for deposition of $HfO_2$ on graphene in this work.

Three different temperature regimes were used for deposition of $HfO_2$ on the as-cleaved graphene samples. First, films with the thicknesses of 30–40 nm were deposited at 170–180 ºC. This temperature range is close to the lowest temperature, at which ALD of $HfO_2$ had been deposited from $HfCl_4$ [16,20]. At somewhat lower temperatures, unlimited condensation of $HfCl_4$ can take place on the substrate surface as the lowest temperatures ensuring sufficient evaporation rates of $HfCl_4$ for ALD range from 130 to 150 ºC [17–22]. Second, 10–30 nm thick films were grown in a two-step process: the growth was initiated at 170 ºC but after applying 10 ALD cycles the temperature was increased to 300 ºC and the growth was completed applying either 100–300 cycles.

As shown earlier, the film growth at lower temperatures should allow more uniform nucleation [21,22] but the concentration of the chlorine and hydrogen residues in the films grown from $HfCl_4$ and $H_2O$ at temperatures below 225 ºC is high, exceeding 2-5% [17,20]. In the two-step process, we tried to achieve uniform nucleation at low temperature in the initial stage of deposition [22] and to grow $HfO_2$ with low impurity concentration and large dielectric constant on the top of the seed layer at a higher temperature. Using this approach, $HfO_2$ layer was also grown on the graphene sample that had electrodes for electrical measurements. For comparison, we deposited $HfO_2$ films on graphene by starting and completing the process at 300 ºC. Those films covered, however, the graphene flakes non-uniformly while the surface roughness exceeded that of the films grown in the two-step process by a factor of 2. For this reason, the films deposited at 300 ºC in a single-step process were not studied in more detail in this work.

After deposition of $HfO_2$ films the samples were again characterized by AFM and Raman spectroscopy techniques. In addition electron probe microanalysis (EPMA) method was applied to estimate the real thickness of $HfO_2$. Using the sample with electrodes, we measured differential AC-conductivity of graphene in a two-lead configuration at a frequency of 32 Hz and temperature of 4.2 K before and after the deposition of an 11 nm thick $HfO_2$ layer in a two-temperature ALD process. The highly doped Si substrate was used as a back gate in these measurements.

## 3. Results and discussion

The image depicted in Fig. 1a confirms that low deposition temperatures yield smooth $HfO_2$ films on graphene as the RMS surface roughness values of about $\delta h_{rms} = 0.5$ nm on graphene and $\delta h_{rms} = 0.35$ nm on $SiO_2$ were measured for a 30 nm thick $HfO_2$ films deposited at 180 ºC. For the films of the same thickness grown in the two-step process (Fig. 1b), these values reached ~2.5 and ~2 nm, respectively. It is worth noting in this context that markedly lower surface roughness ($\delta h_{rms} = 1.7$ nm on graphene and $\delta h_{rms}$ = 0.9 nm on SiO$_2$) was measured for thinner (10–12 nm thick) HfO$_2$ films (Fig. 1c). This result indicates that rougher surfaces of films deposited in the two-step process compared to those of the films deposited at 180 ºC were due to crystallization of HfO$_2$ during the film growth at higher temperature rather than because of nucleation problems on the surface of graphene. At the same time, the surface roughness of 20–30 nm thick films grown at 300 ºC from the beginning of deposition process was as high as 5 nm on graphene and 2 nm on SiO$_2$. Thus, the seed layer deposited at 180 ºC allowed a marked reduction of the surface roughness of HfO$_2$ on graphene but caused no changes in the surface roughness on SiO$_2$. The absence of the effect of the seed layer on SiO$_2$ is an expected result because very uniform nucleation of HfO$_2$ can be obtained on oxidized Si at temperatures up to 300 ºC [21].

At the edge of a monolayer graphene flake, the AFM surface profile of the HfO$_2$ film deposited in the two-step process (Fig. 1d) very well corresponded to that of a single-layer graphene. Thus, the HfO$_2$ film grew with the same rate on graphene and SiO$_2$. The same result was obtained from EPMA studies. Consequently, application of the low-temperature seed layer allows minimization of the incubation period in the beginning of deposition on graphene. One should take into account, however, that although the seed layer obviously resulted in nucleation of the HfO$_2$ films on graphene without marked delay and in relatively smooth films grown in the two-step process, the surface of the films was still somewhat rougher on graphene compared to that on SiO$_2$. Therefore the deposition process parameters and thickness of the seed layer evidently need further optimization.

In order to determine the phase composition of HfO$_2$, we measured Raman scattering of the films on thick graphite flakes because strong scattering coming from Si substrate did not allow reliable characterization of HfO$_2$ on thin graphene flakes. We did not find any Raman peak for HfO$_2$ films deposited at 170–180 ºC. Hence, these films were amorphous. In the films deposited in the two-step growth process, we recorded scattering from monoclinic HfO$_2$ (Fig. 2). This conclusion is based on the comparison of the Raman peaks measured in this work with a reference spectrum measured for a freestanding film that contained monoclinic $HfO_2$ as the only crystalline phase reliably determined by X-ray diffraction analysis [23].

Raman studies performed after deposition of $HfO_2$ indicated also a noticeable shift of the G and 2D bands of graphene compared to the positions of those recorded before $HfO_2$ deposition. Figure 3 shows the Raman spectra of a single-layer graphene sample taken at the same location before and after the deposition of an 11 nm thick $HfO_2$ layer in a two-step process. One can see from Fig. 3 that all Raman peaks of graphene recorded after deposition of $HfO_2$ are weaker than those recorded before $HfO_2$ deposition. In addition, the deposition of $HfO_2$ has resulted in an increase of the background intensity. Finally, and most importantly, the spectra clearly indicate blue-shifts of 9 and 22 $cm^{-1}$ for the Raman G (at ~1580 $cm^{-1}$) and 2D (at ~2670 $cm^{-1}$) peaks, respectively, compared to the peak positions of uncoated graphene. Careful mapping of the whole area of a single-layer graphene (about 5 $\mu m^2$ in this particular sample) showed that the deviations of Raman peak positions did not exceed ±2.5 $cm^{-1}$ for the G peak and ±5 $cm^{-1}$ for the 2D peak. Therefore the blue-shifts observed cannot be because of edge effects [24] and/or possible inaccuracy of the positioning the laser beam at its initial location during Raman measurements. It is known that doping of graphene can also influence the positions of Raman peaks [25,26]. In this case, however, the shift of the 2D band should be smaller than [25] or comparable to [26] that of the G band. In addition, narrowing of the G band should accompany its blue-shift related to edge effects [24] and/or changes in the doping level [26]. In our case, on the contrary, the full width at half maximum of the G peak increases. On the basis of these data, one can conclude that compressive strain developed in graphene is the most probable reason for the blue-shifts of the Raman bands shown in Fig. 3.

Using the biaxial strain coefficient of –58 $cm^{-1}$/% for the Raman G mode and –144 $cm^{-1}$/% for the Raman 2D mode [27] and assuming elastic behavior of graphene, we estimated the compressive strain to be ~0.15% in our single-layer graphene sample. A relatively large and *negative* thermal expansion coefficient of graphene ($-7\times10^{-6}$ K$^{-1}$ at room temperature), which has recently been measured by Bao *et al.* [28], well explains the strain as the thermal expansion of HfO$_2$ is about the same magnitude but with the opposite (positive) sign [29]. We would also like to point out that our ALD process did not introduce additional defects into graphene as there was no measurable rise in the intensity of D (at ~1350 cm$^{-1}$) and D' (at ~1620 cm$^{-1}$) peaks in the Raman spectra.

Figure 4 displays differential conductivity, $\sigma = (L/W)(dI/dV_{ds})$, versus drain-source bias voltage $V_{ds}$ measured at $T = 4.2$ K before and after the two-step growth process of an 11 nm thick HfO$_2$ film. The ALD deposition moved the charge neutrality point (CNP) from –39 V to +30 V in $V_g$ but, nevertheless, the *I-V* characteristics did not change much at CNP (at $V_g^{CNP}$). A comparison of $\sigma$ before and after ALD deposition indicates that the electron mobility $\mu \sim 2000$ cm$^2$/Vs of the sheet was reduced by less than 10% at CNP and by 30–40% far away from it (at $\Delta V_g = V_g - V_g^{CNP} = 23$ V). The electron mobility $\mu$ was estimated from the relationship $\sigma = n e \mu$, where $n$ is the charge carrier density and $e$ is the electron charge. The charge density can be calculated from the formula $n = C_g V_g /e$ where $C_g$ is the gate capacitance per surface area and $V_g$ is the gate voltage. Using the plane capacitance model the value of $C_g$ was estimated to be 180 aF/µm$^2$. In our sample, the spacing between two contacts (300 nm) is close to the SiO$_2$ thickness (250 nm). In that case, the electric field from the back gate is screened by the contacts and the gate capacitance is lower than estimated from the plane capacitance model. Consequently, the mobility of 2000 cm$^2$/Vs, calculated by us, can be underestimated. However, this mobility value is very close to the value observed by Geim et al. [30] for a bilayer graphene.

We also observed that the "Dirac point" moved from $V_g = +30$ V to +10 V in subsequent bias and gate voltage scans. This shift must be related to the loading of charge traps in the ALD coating, since such behavior was much weaker without HfO$_2$. The width of the "conductance dip" at CNP around zero bias (for example, width $\Delta V$ at $\sigma = 10$ $G_0$) is about 30% wider with the ALD coating. Finally, the asymptotic level of conductivity seems to be larger with $HfO_2$ than without it, when $V_g$ is tuned far away from CNP. This might be an indication of improved, annealed contact conditions between graphene and metal electrode as the major part of the ALD process was done at $T = 300$ ºC.

## 4. Conclusions

We have demonstrated in this work that $HfO_2$ films can be successfully grown on non-functionalized graphene by ALD using $HfCl_4$ and $H_2O$ as the precursors. Surface RMS roughness down to 0.5 nm was obtained for 30 nm thick amorphous films deposited on graphene at 180 ºC. By using a uniform, 1 nm thick seed layer produced at 170 ºC, and continuing at a higher temperature (300 ºC), crystalline $HfO_2$ films were grown on graphene. The films obtained were relatively smooth and grew with the same rate on graphene and $SiO_2$. Raman spectroscopy studies of a single-layer graphene revealed no generation of defects in graphene during the ALD process. However, we found that this kind of procedure leads to compressive strain in graphene because of the marked difference in thermal expansion coefficients of graphene and $HfO_2$. The conductivity measurements of a bilayer graphene sample at 4.2 K indicated that the electron mobility was reduced by less than 40% due to the coating of graphene with an $HfO_2$ film in a two-step process. Consequently, the two-step process is applicable for preparation of high-k gate dielectrics for top gates in graphene devices.

**Acknowledgements**

The authors would like to thank M. Paalanen and I. Sildos for fruitful discussions and for granting access to the experimental facilities needed for preparation and characterization of graphene. We are also thankful to R. Danneau and J. Wengler for their contributions at early stages of the work and Jelena Asari for EPMA measurements. This research was carried out with the financial support from Estonian Science Foundation (Grants No. 6651, 6999 and 7845), and Estonian Ministry of Education and Research (targeted project SF0180046s07). One author (HA) also acknowledges the support from European Social fund (Grant MTT1). The work at LTL was supported by the Academy of Finland, EU contract FP6-IST-021285-2 (CARDEQ), and the NANOSYSTEMS project with Nokia Research Center.

**Figure Captions**

**Figure 1.** (color online) AFM images of $HfO_2$ films deposited on top of graphene flakes (a) at 180 ºC, and (b,c) in the two-step (170/300 ºC) growth process. RMS surface roughness values are (a) < 0.5 nm on graphene for a 30 nm thick film, (b) ~2.5 nm on graphene for a 30

nm thick film and (c) 1.7 nm on graphene and 0.9 nm on SiO$_2$ for an 11 nm thick film. (d) AFM height profile of an 11 nm thick HfO$_2$ film grown in the two-step process and measured at the edge of a graphene flake along the line A-A shown in panel (c). Scan areas are 1 x 1 µm$^2$.

**Figure 2.** Raman spectrum of a 30 nm thick monoclinic HfO$_2$ film deposited in the two-step growth process (170/300 °C) on the top of a graphite flake (upper curve). For comparison, a reference spectrum of a free-standing monoclinic HfO$_2$ film is shown by the lower curve. The monoclinic structure of the latter film has been determined by X-ray diffraction analysis [23].

**Figure 3.** Raman spectra of a single-layer graphene sample taken at the same point (a) before and (b) after the deposition of an 11 nm thick HfO$_2$ layer in the two-step (170/300 °C) growth process.

**Figure 4.** (color online) Differential conductivity $\sigma = (L/W) (dI/dV_{ds})$ versus drain-source bias voltage $V_{ds}$ for a bilayer graphene sample (length $L = 0.3$ µm, width $W = 1.3$ µm), measured at $T = 4.2$ K before and after the deposition of an 11 nm thick HfO$_2$ film using the two-step (170/300 °C) growth process; $\Delta V_g$ denotes the back gate voltage offset from the charge neutrality point ("Dirac point").

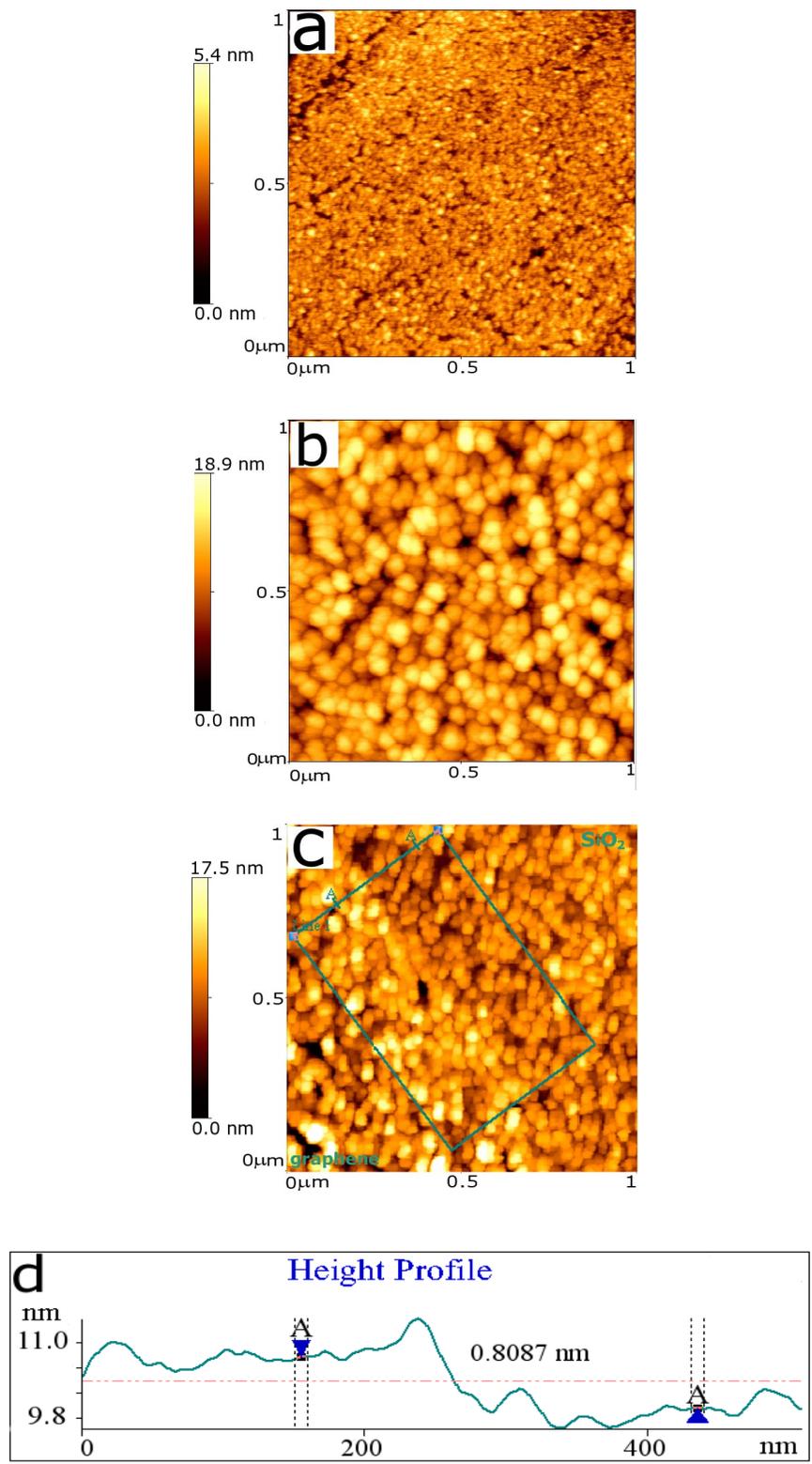

Figure 1

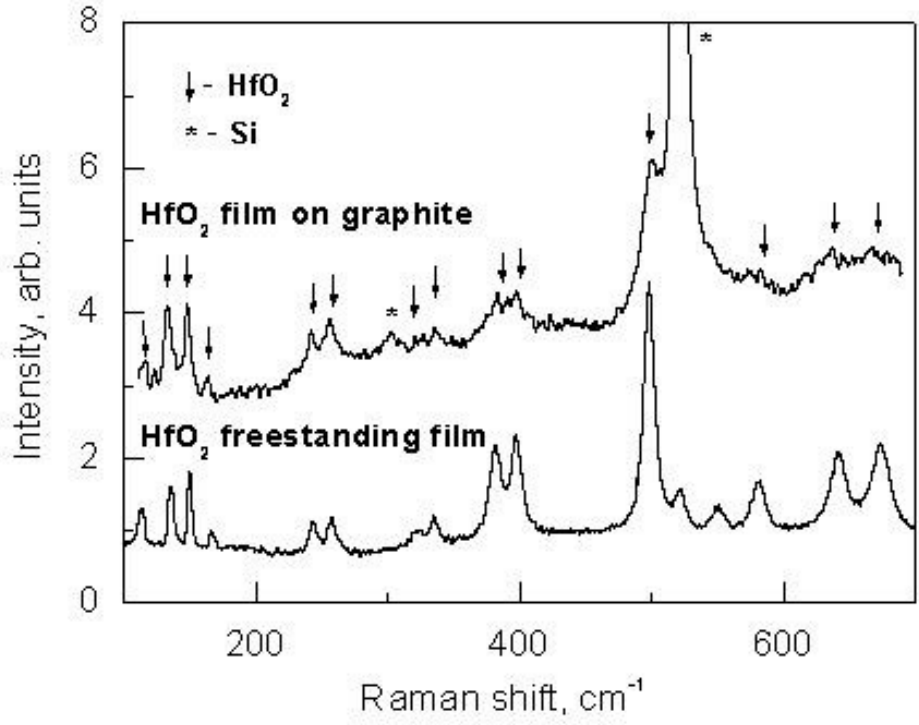

Figure 2

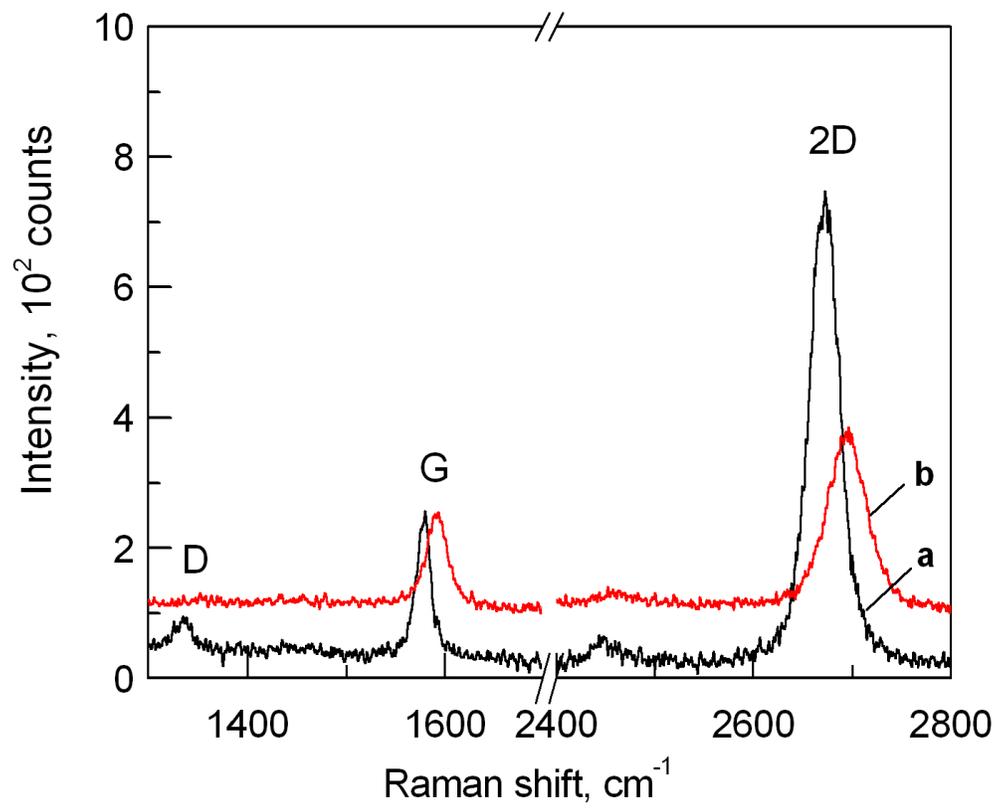

Figure 3

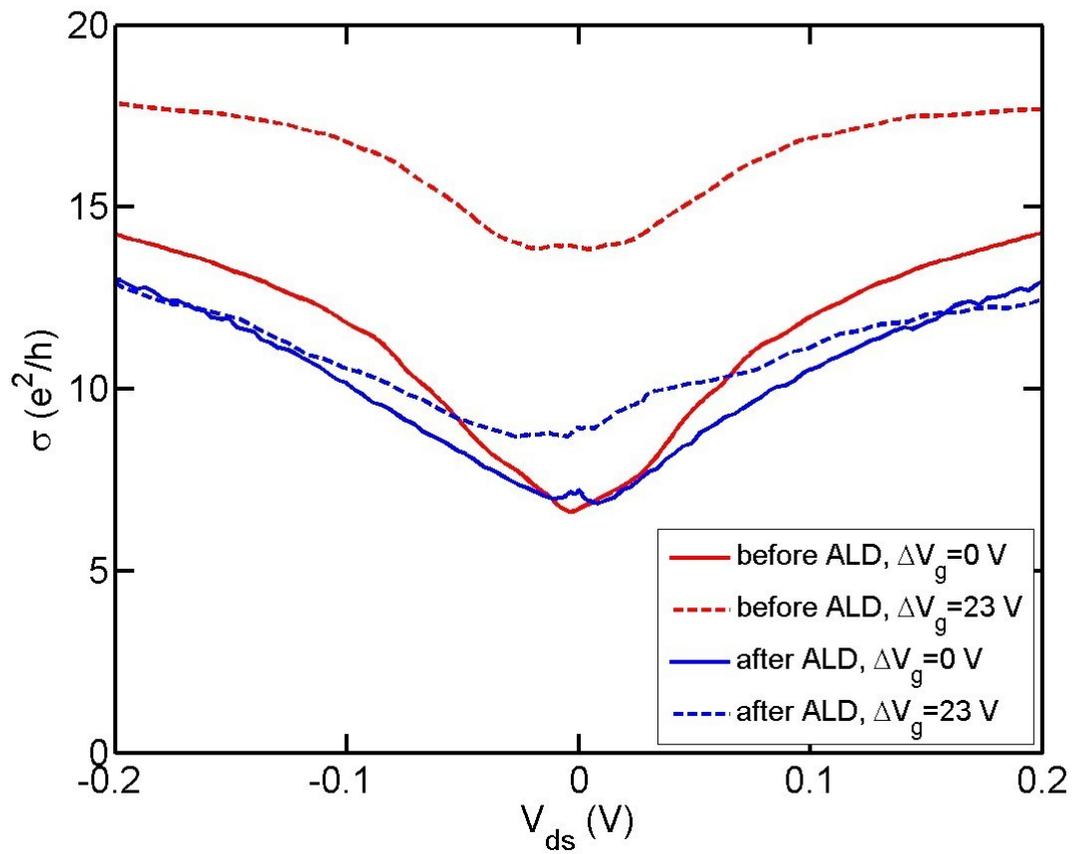

Figure 4